\documentclass[amsmath,notitlepage,amssymb,aps,showkeys,floatfix,prd,a4paper,
  twocolumn,nofootinbib]{revtex4-2}

\usepackage{graphicx}
\usepackage{amsmath}
\usepackage{epstopdf}
\usepackage{amsfonts,amssymb}
\usepackage[utf8]{inputenc}
\usepackage{pstricks}
\usepackage{color}
\usepackage{multirow}
\usepackage{booktabs}
\usepackage{lipsum}

\usepackage[compat=1.0.0]{tikz-feynman}
\usepackage{simplewick}
\usepackage{color, colortbl}
\definecolor{Gray}{gray}{0.9}
\usepackage{float}
\usepackage{tikz-feynman}
\usepackage{feynmp}
\usepackage{forest}

\usepackage{verbatim}    
\usepackage{dcolumn}
\usepackage{bm}
\usepackage{epsfig}
\usepackage{braket}
\usepackage[normalem]{ulem} 
\usepackage{longtable} 

\newcommand{\be}{\begin{equation}}
\newcommand{\ee}{\end{equation}}
\newcommand{\ben}{\begin{eqnarray}}
\newcommand{\een}{\end{eqnarray}}

\def\MeV{\mbox{ MeV}} 
\def\GeV{\mbox{ GeV}}

\def\MeV{\mbox{ MeV}} 
\def\GeV{\mbox{ GeV}} 
\def\mb{\mbox{ mb}} 
\newcommand{\pslash}{\not{\hbox{\kern-2.3pt $p$}}}
\newcommand{\pdslash}{\not{\hbox{\kern-2pt $\partial$}}}

\begin{document}

\title{Hadronic medium effects on $Z_{cs}^- (3985)$ production in heavy
  ion collisions}

\author{ L. M. Abreu}
\email{luciano.abreu@ufba.br}
\affiliation{ Instituto de F\'isica, Universidade Federal da Bahia,
Campus Universit\'ario de Ondina, 40170-115, Bahia, Brazil}

\author{F. S. Navarra}
\email{navarra@if.usp.br}
\affiliation{Instituto de F\'{\i}sica, Universidade de S\~{a}o Paulo, 
Rua do Mat\~ao, 1371, CEP 05508-090,  S\~{a}o Paulo, SP, Brazil}

\author{H. P. L. Vieira}
\email{hildeson.paulo@ufba.br}
\affiliation{ Instituto de F\'isica, Universidade Federal da Bahia,
Campus Universit\'ario de Ondina, 40170-115, Bahia, Brazil}

\begin{abstract}

In this work we study the interactions of the multiquark 
state $Z_{cs}^-(3985)$ with light mesons in a hot hadron gas.        
Using an effective Lagrangian framework, we estimate the vacuum cross
sections as well as the thermal  cross sections of the production  
processes $ \bar{D}_{s}^{(*)}  D_{s}^{(*)} \rightarrow Z_{cs}^-  X \, 
(X=\pi , K, \eta) $ and the corresponding inverse reactions.
The results indicate that the considered processes have sizeable cross
sections.
Most importantly, the thermal cross sections for  $Z_{cs}$ annihilation are
much larger than those for production.  This feature might produce relevant 
effects on some observables, such as the final $Z_{cs}$ multiplicity
measured in heavy ion collisions.

\end{abstract}

\maketitle


\section{Introduction}

\label{Introduction} 

Recently, the BES-III Collaboration has observed an excess of events in
the $K^+$ recoil-mass spectrum of the reaction
$e^+ e^- \rightarrow K^+ (D_{s}^{* -} D^{0} + D_{s}^{-} D^{* 0} ) $ 
for events collected at center-of-mass energy $\sqrt{s} = 4.681 \GeV$, with  
estimated statistical significance of $5.3 \, \sigma$~\cite{BESIII:2020qkh}.
By using an amplitude model based on the Breit-Wigner formalism, this peak has
been fitted to a resonance with mass and width given by
$M = (3982.5 _{-2.6}^{+1.8} \pm 2.1) \MeV, \,\, \Gamma = (12.8 _{-4.4}^{+5.3}
\pm 3.0) \MeV $,
respectively, and has been denoted as  $Z_{cs}^-(3985)$. 
Its minimum valence quark content should be most likely $c \bar{c} s \bar{u}  $,
giving it the status of the first candidate for a charged hidden-charm
tetraquark with strangeness.  

Since the experimental discovery of the $Z_{cs}^-(3985)$ state (or simply   
$Z_{cs}^-$), the  hadron spectroscopy community has been intensely debating 
its internal structure and the possible mechanisms of its decay and
production~\cite{lnw-2009,Yang:2020nrt,Meng:2020ihj,Sun:2020hjw,
  Wang:2020htx,Xu:2020evn,
Yan:2021tcp,Ortega:2021enc,Wang:2020iqt,Jin:2020yjn,Ortega:2021sdd,
Garcilazo:2021nyz,Ikeno:2020mra,Llanes-Estrada:2021jud,Du:2022jjv,
Wang:2020kej,Chen:2020yvq,Azizi:2020zyq,Sungu:2020zvk,Liu:2020nge,Wu:2021ezz,
Ferretti:2021zis,Ikeno:2021mcb,Wu:2021cyc,Chen:2022yev,Han:2022yst}.
Because of its proximity to the  $ D_{s}^{* -} D^{0} $ and $ D_{s}^{-} D^{* 0}$
thresholds, the hadronic molecular interpretation for the $Z_{cs}^-$ seems
natural. Along this line,  this new state would be the strange partner of the
$Z_c(3900)$~\cite{Yang:2020nrt,Meng:2020ihj,Sun:2020hjw,Wang:2020htx,Xu:2020evn, 
  Yan:2021tcp,Ortega:2021enc}. Notwithstanding,  other possible interpretations
have also been proposed, namely: the compact tetraquark configuration resulting 
from the binding of a diquark and an antidiquark    
~\cite{Wang:2020iqt,Jin:2020yjn,Garcilazo:2021nyz}; a virtual pole 
state~\cite{Ortega:2021sdd}; a  kinematic effect caused by triangle 
singularities~\cite{Ikeno:2020mra,Llanes-Estrada:2021jud}; a              
resonance~\cite{Du:2022jjv}, and so on. More experimental and theoretical
studies are clearly needed. 

A new and promising scenario to investigate the properties of exotic states are 
heavy-ion collisions (HICs). They are characterized by the formation of a 
locally thermalized state of deconfined quarks and gluons
(the quark-gluon plasma or QGP). At the end of the QGP phase, quarks
coalesce to form  conventional bound states and also exotic states. The latter
will exist in a hadron gas and interact with other light hadrons. As pointed
out in previous studies, the exotic states can be destroyed in collisions with
the comoving light mesons, as well as produced through the inverse processes
\cite{ChoLee1,XProd1,XProd2,Abreu:2017cof,Abreu:2018mnc,Hong:2018mpk,      
  Abreu:2021jwm,Abreu:2022lfy}. Their final yields depend on the interaction 
cross sections, which, in turn, depend on the spatial configuration of the   
quarks. In the study of the most famous exotic state, the $X(3872)$,  it has
been shown that  the molecular configuration
(i.e., the bound state $(D \bar{D}^* + c.c.)$) is larger than a
diquark-antidiquark configuration $[(c q)(\bar{c}\bar{q})]$ by 
a factor about 3-10~\cite{XProd2}. Consequently, meson molecules have larger 
cross sections and are expected to be more easily produced as well as more
easily destroyed than compact tetraquarks in a hadronic medium. 

The recent observation of the $X(3872)$ in $Pb-Pb$ collisions at
$\sqrt{s_{NN}} = 5.02$ TeV by the CMS Collaboration~\cite{CMS:2021znk} 
has opened a new era for the study of exotic states. This observation
strengthens our belief 
that  HICs provide an unique and promising experimental environment to
study the nature of exotic hadrons.

The present contribution is part of a series of works devoted to the 
production of exotics states in heavy ion collisions. In the 
following sections we will analyze the interactions of the
$Z_{cs}^-$  state with light mesons. 
In  Section~\ref{formalism} we present our effective Lagrangian formalism.
In Section ~\ref{Cross Sections} we use it to calculate the 
$Z_{cs}^-$ production and absorption cross sections and in 
Section~\ref{ThermalAvCrossSection} we compute the corresponding thermal  
averages.  Finally, Section~\ref{Conclusions} is dedicated to the summary
and to the concluding remarks.

\section{ The formalism }

\label{formalism} 


\begin{figure}[!ht]
    \centering

\begin{tikzpicture}
\begin{feynman}
\vertex (a1) {$\bar{D}_s (p_1)$};
	\vertex[right=1.5cm of a1] (a2);
	\vertex[right=1.cm of a2] (a3) {$Z_{cs} (p_3)$};
	\vertex[right=1.4cm of a3] (a4) {$\bar{D}^{*}_{s} (p_1)$};
	\vertex[right=1.5cm of a4] (a5);
	\vertex[right=1.cm of a5] (a6) {$Z_{cs} (p_{3})$};
\vertex[below=1.5cm of a1] (c1) {$D (p_2)$};
\vertex[below=1.5cm of a2] (c2);
\vertex[below=1.5cm of a3] (c3) {$\pi (p_4)$};
\vertex[below=1.5cm of a4] (c4) {$D^{*} (p_2)$};
\vertex[below=1.5cm of a5] (c5);
\vertex[below=1.5cm of a6] (c6) {$\pi (p_4)$};
	\vertex[below=2cm of a2] (d2) {(a)};
	\vertex[below=2cm of a5] (d5) {(b)};
\diagram* {
(a1) -- (a2), (a2) -- (a3), (c1) -- (c2), (c2) -- (c3), (c2) -- [fermion, edge label'= $D^{*}$] (a2), (a4) -- (a5), (a5) -- (a6), (c4) -- (c5), (c5) -- (c6), (c5) -- [fermion, edge label'= $D$] (a5)
}; 
\end{feynman}
\end{tikzpicture}

\begin{tikzpicture}
\begin{feynman}
\vertex (a1) {$\bar{D}_s (p_1)$};
	\vertex[right=1.5cm of a1] (a2);
	\vertex[right=1.cm of a2] (a3) {$Z_{cs} (p_3)$};
	\vertex[right=1.4cm of a3] (a4) {$D (p_1)$};
	\vertex[right=1.5cm of a4] (a5);
	\vertex[right=1.cm of a5] (a6) {$Z_{cs} (p_{3})$};
\vertex[below=1.5cm of a1] (c1) {$D^* (p_2)$};
\vertex[below=1.5cm of a2] (c2);
\vertex[below=1.5cm of a3] (c3) {$\pi (p_4)$};
\vertex[below=1.5cm of a4] (c4) {$\bar{D} (p_2)$};
\vertex[below=1.5cm of a5] (c5);
\vertex[below=1.5cm of a6] (c6) {$K (p_4)$};
	\vertex[below=2cm of a2] (d2) {(c)};
	\vertex[below=2cm of a5] (d5) {(d)};
\diagram* {
(a1) -- (a2), (a2) -- (a3), (c1) -- (c2), (c2) -- (c3), (c2) -- [fermion, edge label'= $D^{*}$] (a2), (a4) -- (a5), (a5) -- (a6), (c4) -- (c5), (c5) -- (c6), (c5) -- [fermion, edge label'= $D_{s}^{*}$] (a5)
}; 
\end{feynman}
\end{tikzpicture}

\begin{tikzpicture}
\begin{feynman}
\vertex (a1) {$D^{*} (p_1)$};
	\vertex[right=1.5cm of a1] (a2);
	\vertex[right=1.cm of a2] (a3) {$Z_{cs} (p_3)$};
	\vertex[right=1.4cm of a3] (a4) {$D (p_1)$};
	\vertex[right=1.5cm of a4] (a5);
	\vertex[right=1.cm of a5] (a6) {$Z_{cs} (p_{3})$};
\vertex[below=1.5cm of a1] (c1) {$\bar{D}^{*} (p_2)$};
\vertex[below=1.5cm of a2] (c2);
\vertex[below=1.5cm of a3] (c3) {$K (p_4)$};
\vertex[below=1.5cm of a4] (c4) {$\bar{D}^* (p_2)$};
\vertex[below=1.5cm of a5] (c5);
\vertex[below=1.5cm of a6] (c6) {$K (p_4)$};
	\vertex[below=2cm of a2] (d2) {(e)};
	\vertex[below=2cm of a5] (d5) {(f)};
\diagram* {
(a1) -- (a2), (a2) -- (a3), (c1) -- (c2), (c2) -- (c3), (c2) -- [fermion, edge label'= $\bar{D}_{s}$] (a2), (a4) -- (a5), (a5) -- (a6), (c4) -- (c5), (c5) -- (c6), (c5) -- [fermion, edge label'= $\bar{D}_{s}^{*}$] (a5)
}; 
\end{feynman}
\end{tikzpicture}

\begin{tikzpicture}
\begin{feynman}
\vertex (a1) {$\bar{D}_s (p_1)$};
	\vertex[right=1.5cm of a1] (a2);
	\vertex[right=1.cm of a2] (a3) {$Z_{cs} (p_3)$};
	\vertex[right=1.4cm of a3] (a4) {$\bar{D}_s (p_1)$};
	\vertex[right=1.5cm of a4] (a5);
	\vertex[right=1.cm of a5] (a6) {$Z_{cs} (p_{3})$};
\vertex[below=1.5cm of a1] (c1) {$D_s (p_2)$};
\vertex[below=1.5cm of a2] (c2);
\vertex[below=1.5cm of a3] (c3) {$K (p_4)$};
\vertex[below=1.5cm of a4] (c4) {$D_s^{*} (p_2)$};
\vertex[below=1.5cm of a5] (c5);
\vertex[below=1.5cm of a6] (c6) {$K (p_4)$};
	\vertex[below=2cm of a2] (d2) {(g)};
	\vertex[below=2cm of a5] (d5) {(h)};
\diagram* {
(a1) -- (a2), (a2) -- (a3), (c1) -- (c2), (c2) -- (c3), (c2) -- [fermion, edge label'= $D^{*}$] (a2), (a4) -- (a5), (a5) -- (a6), (c4) -- (c5), (c5) -- (c6), (c5) -- [fermion, edge label'= $D^{*}$] (a5)
}; 
\end{feynman}
\end{tikzpicture}

\begin{tikzpicture}
\begin{feynman}
\vertex (a1) {$\bar{D}_s (p_1)$};
	\vertex[right=1.5cm of a1] (a2);
	\vertex[right=1.cm of a2] (a3) {$Z_{cs} (p_3)$};
	\vertex[right=1.4cm of a3] (a4) {$\bar{D}_s (p_1)$};
	\vertex[right=1.5cm of a4] (a5);
	\vertex[right=1.cm of a5] (a6) {$Z_{cs} (p_{3})$};
\vertex[below=1.5cm of a1] (c1) {$D (p_2)$};
\vertex[below=1.5cm of a2] (c2);
\vertex[below=1.5cm of a3] (c3) {$\eta (p_4)$};
\vertex[below=1.5cm of a4] (c4) {$D (p_2)$};
\vertex[below=1.5cm of a5] (c5);
\vertex[below=1.5cm of a6] (c6) {$\eta (p_4)$};
	\vertex[below=2cm of a2] (d2) {(i)};
	\vertex[below=2cm of a5] (d5) {(j)};
\diagram* {
(a1) -- (a2), (a2) -- (a3), (c1) -- (c2), (c2) -- (c3), (c2) -- [fermion, edge label'= $D^{*}$] (a2), (a4) -- (a5), (a5) -- (c6), (c4) -- (c5), (c5) -- (a6), (a5) -- [fermion, edge label'= $\bar{D}_{s}^{*}$] (c5)
}; 
\end{feynman}
\end{tikzpicture}

\begin{tikzpicture}
\begin{feynman}
\vertex (a1) {$\bar{D}_{s}^{*} (p_1)$};
	\vertex[right=1.5cm of a1] (a2);
	\vertex[right=1.cm of a2] (a3) {$Z_{cs} (p_3)$};
	\vertex[right=1.4cm of a3] (a4) {$\bar{D}_{s}^{*} (p_1)$};
	\vertex[right=1.5cm of a4] (a5);
	\vertex[right=1.cm of a5] (a6) {$Z_{cs} (p_{3})$};
\vertex[below=1.5cm of a1] (c1) {$D^{*} (p_2)$};
\vertex[below=1.5cm of a2] (c2);
\vertex[below=1.5cm of a3] (c3) {$\eta (p_4)$};
\vertex[below=1.5cm of a4] (c4) {$D^{*} (p_2)$};
\vertex[below=1.5cm of a5] (c5);
\vertex[below=1.5cm of a6] (c6) {$\eta (p_4)$};
	\vertex[below=2cm of a2] (d2) {(k)};
	\vertex[below=2cm of a5] (d5) {(l)};
\diagram* {
(a1) -- (a2), (a2) -- (a3), (c1) -- (c2), (c2) -- (c3), (c2) -- [fermion, edge label'= $D$] (a2), (a4) -- (a5), (a5) -- (c6), (c4) -- (c5), (c5) -- (a6), (a5) -- [fermion, edge label'= $\bar{D}_{s}$] (c5)
}; 
\end{feynman}
\end{tikzpicture}

\begin{tikzpicture}
\begin{feynman}
\vertex (a1) {$\bar{D}_{s}^{*} (p_1)$};
	\vertex[right=1.5cm of a1] (a2);
	\vertex[right=1.cm of a2] (a3) {$Z_{cs} (p_3)$};
\vertex[below=1.5cm of a1] (c1) {$D (p_2)$};
\vertex[below=1.5cm of a2] (c2);
\vertex[below=1.5cm of a3] (c3) {$\eta (p_4)$};
	\vertex[below=2cm of a2] (d2) {(m)};
\diagram* {
(a1) -- (a2), (a2) -- (c3), (c1) -- (c2), (c2) -- (a3), (a2) -- [fermion, edge label'= $\bar{D}_{s}^{*}$] (c2)
}; 
\end{feynman}
\end{tikzpicture}
\caption{Diagrams contributing to the following process (without specification 
  of the charges of the particles): $ \bar{D}_{s}^{(*)} D^{(*)}
  \rightarrow Z_{cs} \pi $  [(a)-(c)], $ D^{(*)} \bar{D}^{(*)} , \bar{D}_{s}
  D_s^{(*)}  \rightarrow Z_{cs} K $ [(d)-(h)] and $ \bar{D}_{s}^{(*)} D^{(*)}
  \rightarrow Z_{cs}\eta $  [(i)-(m)]. The particle charges are not specified. }
\label{DIAG1}
\end{figure}
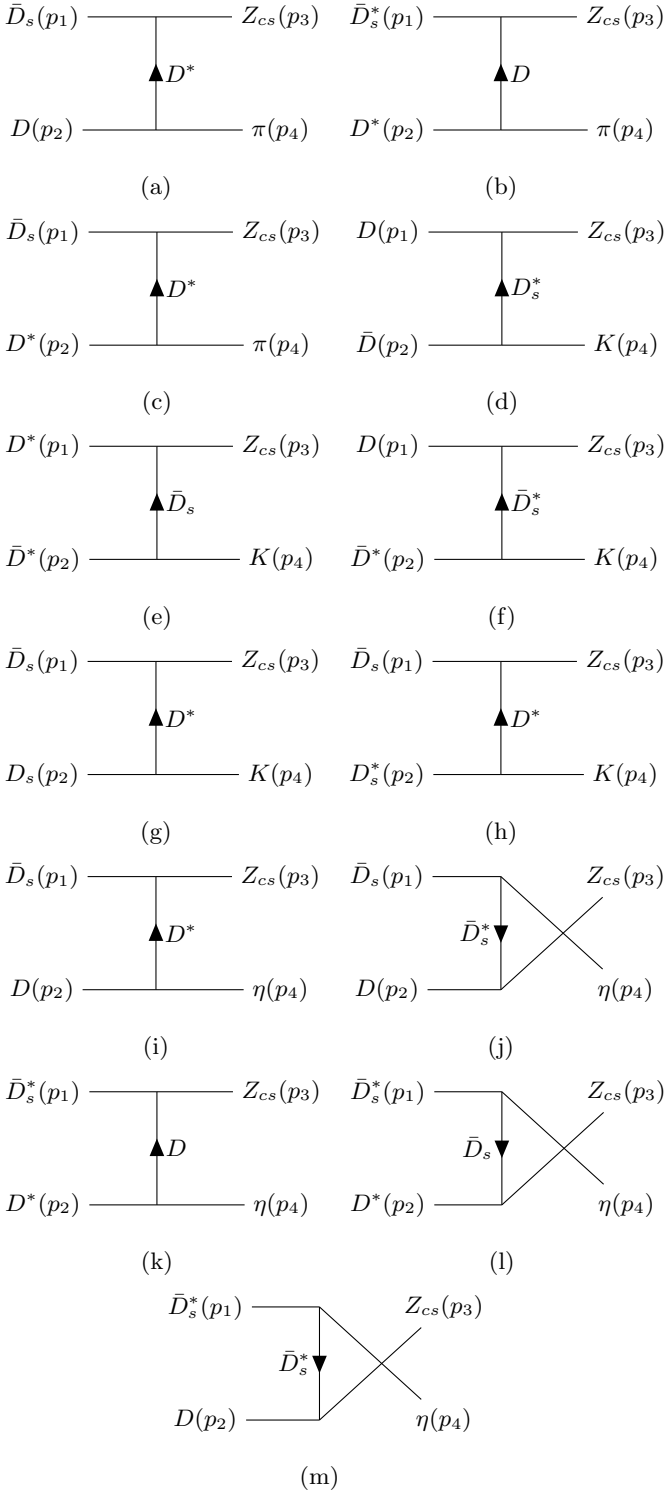



To understand how the $Z_{cs}^-$ behaves in a surrounding hadronic medium, 
we will study its interactions with the lightest pseudoscalar mesons $\pi$,
$K$ and $\eta$. More precisely, we will focus on  the reactions 
$  \bar{D}_{s}^{(*)} D^{(*)}  \rightarrow Z_{cs} \pi $, $  D^{(*)} 
\bar{D}^{(*)} , \bar{D}_{s} D_s^{(*)} \rightarrow Z_{cs} K $  and    
$ \bar{D}_{s}^{(*)} D^{(*)} \rightarrow Z_{cs}\eta $, as well as the
inverse processes. In Fig.~\ref{DIAG1} we present the   
lowest-order Born diagrams contributing to these processes, without
specifying the charge of the particles. 


In the evaluation of the reactions in Fig.~\ref{DIAG1}, we make use of the
effective theory approach. Consequently, the couplings involving    
$\pi$, $K^{(*)}$, $D^{(*)}$ and $D_s^{(*)}$ mesons are based on the        
effective formalism in which the vector mesons are identified as dynamical     
gauge bosons of the hidden $U(N)_V$ local symmetry, and are properly explained
in Refs.~\cite{ChoLee1,XProd1,XProd2,Abreu:2017cof,Abreu:2018mnc}; they read
\begin{eqnarray}\label{Lagr1}
\mathcal{L}_{\pi D D^*} &=& i g_{\pi DD^*} D^{*}_{\mu} \vec{\tau}\cdot (\bar{D} \partial^{\mu} \bar{\pi} - \partial^{\mu} \bar{D} \bar{\pi}) + h.c., \nonumber  \\
\mathcal{L}_{K D_s D^*} &=&i g_{K D_s D^*} D^{*+}_{\mu} (K \partial^{\mu} D_s - \partial^{\mu} K D_s) + h.c.,\nonumber  \\
\mathcal{L}_{KDD^{*}_{s}} &=& i g_{KDD^{*}_{s}} D^{*\mu}_{s} (\partial_{\mu} D^+ K - D^+ \partial_{\mu} K) + h.c., \nonumber  \\
\mathcal{L}_{\eta DD^*} &=& i g_{\eta DD^*}  D^{*}_{\mu} (D^+ \partial^{\mu}\eta - \partial^{\mu} D^+) + h.c., \nonumber   \\
\mathcal{L}_{\eta D_s D^{*}_{s}} &=& i g_{\eta D_s D^{*}_{s}}  D^{*}_{s\mu} (\partial^{\mu} D^{+}_{s} \eta - \partial^{\mu}_{\eta} D^{+}_{s}) + h.c., \nonumber   \\
\mathcal{L}_{\pi D^* D^*} &=& -g_{\pi D^* D^*} \varepsilon^{\mu \gamma \alpha \beta} \partial_{\mu} D^{*}_{\nu} \vec{\pi} \partial_{\alpha} D^{*+}_{\beta}, \nonumber  \\
\mathcal{L}_{\eta D^* D^*} &=& -g_{\eta D^* D^*} \varepsilon^{\mu \nu \alpha \beta} \partial_{\mu} D^{*}_{\nu} \eta \partial_{\alpha} D^{*+}_{\beta},\nonumber   \\
\mathcal{L}_{\eta D^{*}_{s} D^{*}_{s}} &=& -g_{\eta D^{*}_{s} D^{*}_{s}} \varepsilon^{\mu \nu \alpha \beta} \partial_{\mu} D^{*}_{s\nu} \eta \partial_{\alpha} D^{*}_{s\beta}, \nonumber  \\
\mathcal{L}_{K D^{*}_{s} D^{*}} &=& g_{ K D^{*}_{s} D^*}  \varepsilon^{\mu \nu \alpha \beta} \partial_{\mu} D^{*+}_{\nu} K \partial_{\alpha} D^{*}_{s \beta} + h.c. , 
\end{eqnarray}
where $\vec{\tau}$ are the Pauli matrices in the isospin space;         
$\vec{\pi}$ denotes the pion isospin triplet; and
$D^{(\ast)} = (D^{(\ast) 0}, D^{(\ast) +}  ) $ and $K = (K ^{+}, K^{ 0}  )^T $
represent the isospin doublets for the pseudoscalar (vector) $D^{(\ast) }$
and $K$ mesons, respectively. 

The coupling constants in Eq.~(\ref{Lagr1}) describe
pseudoscalar-pseudoscalar-vector and vector-vector-pseudoscalar vertices 
and are given by~\cite{XProd1,XProd2,Abreu:2017cof,Abreu:2018mnc},
\begin{eqnarray}
g_{\pi DD^*} & = &  g_{K D_s D^*} = g_{KDD^{*}_{s}}  \nonumber   \\ 
& = &  \sqrt{6} g_{\eta DD^*}  = \sqrt{3} g_{\eta D_s D^{*}_{s}} \equiv g_{PPV};
\nonumber   \\
\sqrt{2}g_{\pi D^* D^*} & = & 2\sqrt{3} g_{\eta D^* D^*}  =
\sqrt{3} g_{\eta D^{*}_{s} D^{*}_{s}}  \nonumber   \\ 
& = &  \sqrt{2} g_{ K D^{*}_{s} D^*} \equiv g_{VVP}, 
\label{eq:CouplingConstants1}
\end{eqnarray} 
where 
\begin{eqnarray}
 g_{PPV} & = & \frac{m_V}{2f_\pi} \frac{m_{D^*}}{m_{K^*}}, \nonumber   \\
g_{VVP} &  =  &\frac{3m^2_V}{16\pi^2f^3_\pi}, 
\label{eq:CouplingConstants2}
\end{eqnarray}
with $m_V$ being the mass of  the  vector meson; we take it as the mass of
the $\rho$ meson and $f_\pi$ is the pion decay constant. As pointed in      
Ref.~\cite{XProd1}, the factor $m_{D^*}/m_{K^*} $ in the coupling $g_{PPV}$ 
is introduced in order to reproduce the experimental decay width found for
the process $D^* \to D\pi$, and comes from heavy-quark symmetry considerations.

The couplings involving the $Z_{cs}^-$ are introduced assuming that it is a    
$S$-wave bound state engendered by the superposition of $  D_{s}^{* -} D^{0}$ 
and $  D_{s}^{-} D^{* 0} $  configurations with quantum numbers            
$I (J^P) = \frac{1}{2} (1^+) $. As a consequence, the effective Lagrangian  
describing the interaction between the  $Z_{cs}^-$ and the
$  D_{s}^{* -} D^{0} $ and $  D_{s}^{-} D^{* 0}$
pairs is given by~\cite{Wu:2021ezz},   
\begin{eqnarray}\label{Lagr3}
  \mathcal{L}_{Z_{cs}} = \frac{ g_{Z_{cs}} }{ \sqrt{2} } Z_{cs}^{\dagger \mu}
  ( \bar{D}_{s \mu}^{*} D + \bar{D}_{s} D_{\mu}^{*} ),
\end{eqnarray}
where  $Z_{cs}$ denotes the field associated to  
$Z_{cs}^-$ state; this notation will be used henceforth. Also, the       
$\bar{D}_{s \mu}^{*} D $ and $\bar{D}_{s} D_{\mu}^{*}$ mean the
$D_{s}^{* -} D^{0}$ and  $D_{s}^{-} D^{* 0}$ 
components, respectively.  The effective coupling constant $ g_{Z_{cs}} $
is considered to be $g_{Z_{cs}} = 6.0 - 6.7$ in order to describe the $Z_{cs} $
width, as discussed in Ref.~\cite{Wu:2021ezz}.

%

Based on the effective Lagrangians introduced above, the amplitudes of the processes shown in Fig.~\ref{DIAG1} can then be calculated. They are given by
\begin{eqnarray}
	\nonumber \mathcal{M}_{\bar{D}_{s} D \rightarrow Z_{cs}\pi} &\equiv &  \mathcal{M}^{(a)},\\
	\nonumber \mathcal{M}_{\bar{D}^{*}_{s}D^* \rightarrow Z_{cs}\pi}  &\equiv & \mathcal{M}^{(b)}, \\
	\nonumber \mathcal{M}_{\bar{D}_{s} D^* \rightarrow Z_{cs}\pi}  &\equiv & \mathcal{M}^{(c)}, \\
	\nonumber \mathcal{M}_{D\bar{D} \rightarrow Z_{cs}K}  &\equiv &\mathcal{M}^{(d)},\\
	\nonumber \mathcal{M}_{D^{*}\bar{D}^{*} \rightarrow Z_{cs}K}  &\equiv & \mathcal{M}^{(e)},\\
	\nonumber \mathcal{M}_{D\bar{D}^{*} \rightarrow Z_{cs}K}  &\equiv & \mathcal{M}^{(f)},\\
	\nonumber \mathcal{M}_{\bar{D}_{s}D_{s} \rightarrow Z_{cs}K}  &\equiv & \mathcal{M}^{(g)},\\
	\nonumber \mathcal{M}_{\bar{D}_{s}D^{*}_{s} \rightarrow Z_{cs}K}  &\equiv & \mathcal{M}^{(h)},\\
	\nonumber \mathcal{M}_{\bar{D}_{s} D \rightarrow Z_{cs}\eta}  &\equiv & \mathcal{M}^{ (i) } + \mathcal{M}^{ (j) }, \\
	\nonumber \mathcal{M}_{\bar{D}^{*}_{s}D^* \rightarrow Z_{cs}\eta}  &\equiv & \mathcal{M}^{ (k) } + \mathcal{M}^{ (l) },\\		\
	\mathcal{M}_{\bar{D}^{*}_{s} D \rightarrow Z_{cs}\eta}  &\equiv & \mathcal{M}^{ (m) },
\label{Ampl1}
\end{eqnarray}
where the explicit expressions are 
\begin{eqnarray}
	\nonumber \mathcal{M}^{(a)} & = & \frac{1}{\sqrt{2}} \tau _{I}  g_{Z_{cs}} g_{\pi D D^*} (p_2 + p_4)^{\mu} \varepsilon^{\nu}_{3} \frac{1}{t - m^2_{D^*}} \\
	& & \times \left[ -g_{\mu \nu} + \frac{(p_1 - p_3)_{\mu} (p_1 - p_3)_{\nu}}{m^{2}_{D^*}} \right] ,
\nonumber \\
	\mathcal{M}^{(b)} & = & -\frac{1}{\sqrt{2}} \tau _{I}  g_{Z_{cs}} g_{\pi D D^*} \varepsilon^{\mu}_{2} (2p_4 - p_2)_{\mu} \varepsilon^{\nu}_{1} \varepsilon^{*}_{3\nu} \frac{1}{t - m^2_D},
\nonumber \\
	\mathcal{M}^{ (c) } & = & \frac{1}{\sqrt{2}} \tau _{I}  g_{Z_{cs}} g_{\pi D^* D^*} \epsilon^{\mu \nu \alpha \beta} p_{2\mu} p_{4\alpha} \varepsilon_{2\nu} \varepsilon^{*}_{3\beta} \frac{1}{t - m^{2}_{D^*}},\nonumber \\
		 \mathcal{M}^{ (d) } & = & \frac{1}{\sqrt{2}} \tau_{ij} g_{Z_{cs}} g_{K D D^{*}_{s}} (p_2 + p_4)^{\mu} \varepsilon^{*\nu}_{3} \frac{1}{t - m^{2}_{D^*_s}} \nonumber \\
	& & \times \left[ -g_{\mu \nu} + \frac{(p_1 - p_3)_\mu (p_1 - p_3)_{\nu}}{m^{2}_{D^*_s}} \right] ,
\nonumber \\
	\mathcal{M}^{ (e) } & = & \frac{1}{\sqrt{2}} \tau _{I}  g_{Z_{cs}} g_{K D_s D^*} \varepsilon^{\mu}_{2} (2p_4 - p_2)_{\mu} \varepsilon^{\nu}_{1} \varepsilon^{*}_{3\nu} \frac{1}{t - m^2_{D_s}} ,
\nonumber \\
	\mathcal{M}^{ (f) } & = & \frac{1}{\sqrt{2}} \tau _{I}  g_{Z_{cs}} g_{K D^{*}_{s} D^*} \epsilon^{\mu \nu \alpha \beta} p_{2\mu} p_{4\alpha} \varepsilon_{2\nu} \varepsilon^{*}_{3\beta} \frac{1}{t - m^{2}_{D^*_s}},
\nonumber \\
 \mathcal{M}^{ (g) } & = & \frac{1}{\sqrt{2}} \tau _{I}  g_{Z_{cs}} g_{k D_s D^*} (p_2 + p_4)^{\mu} \varepsilon^{\nu}_{3} \frac{1}{t - m^2_{D^*}} \nonumber \\
	& & \times\left( -g_{\mu \nu} + \frac{(p_1 - p_3)_{\mu} (p_1 - p_3)_{\nu}}{m^{2}_{D^*}} \right) ,	
\label{Ampl2a}
\end{eqnarray}
and
\begin{eqnarray}	\nonumber	\mathcal{M}^{ (h) } & = & \frac{1}{\sqrt{2}} \tau _{I}  g_{Z_{cs}} g_{k D^*_s D^*} \epsilon^{\mu \nu \alpha \beta } p_{2\mu} p_{4\alpha} \varepsilon_{2\nu} \varepsilon^{*}_{3\beta} \frac{1}{t - m^{2}_{D^*}},
\nonumber \\
	\nonumber \mathcal{M}^{ (i) } & = & \frac{1}{\sqrt{2}} \tau _{I}  g_{Z_{cs}} g_{\eta D^* D^*} (p_2 + p_4)^{\mu} \varepsilon^{\nu}_{3} \frac{1}{t - m^{2}_{D^*}} \\
	& & \times \left[ -g_{\mu \nu} +  \frac{(p_1 - p_3)_{\mu} (p_1 - p_3)_{\nu}}{m^{2}_{D^*}} \right] ,
\nonumber \\
	\nonumber \mathcal{M}^{ (j) } & = & -\frac{1}{\sqrt{2}} \tau _{I}  g_{Z_{cs}} g_{\eta D_s D^{*}_{s}} \frac{1}{ u - m^{2}_{D^*_s}} \varepsilon^{*\mu}_{3} \\
	& & \times \left[ -g_{\mu \nu} +  \frac{(p_1 - p_4)_{\mu} (p_1 - p_4)_{\nu}}{m^{2}_{D^*_s}} \right] (p_2 + p_3)^{\nu}, 
\nonumber \\
	\mathcal{M}^{ (k) } & = & \frac{1}{\sqrt{2}} \tau _{I}  g_{Z_{cs}} g_{\eta D D^*} \varepsilon^{\mu}_{2} (2p_2 - p_4)_{\mu} \varepsilon^{\nu}_{1} \varepsilon^{*}_{3\nu} \frac{1}{t - m^{2}_{D}},
\nonumber \\
	\mathcal{M}^{ (l) } & = & \frac{1}{\sqrt{2}} \tau _{I}  g_{Z_{cs}} g_{\eta D_s D^{*}_{s}} \varepsilon^{*\mu}_{3} \varepsilon^{\nu}_{1} \varepsilon_{2\mu} \frac{1}{u-m^{2}_{D_s}} (2p_4 - p_1)_{\nu},
\nonumber \\
	\mathcal{M}^{ (m) } & = & \frac{1}{\sqrt{2}} \tau _{I}  g_{Z_{cs}} g_{\eta D^*_s D^*_s} \epsilon^{\mu \nu \alpha \beta } p_{1\mu} \varepsilon_{1\nu} p_{4 \alpha} \varepsilon^{*}_{3\beta} \frac{1}{u - m^{2}_{D^*_s}}.
	\nonumber \\
\label{Ampl2b}
\end{eqnarray}
where $\tau _{I} $ is the isospin factor related to part of the particles in the vertices $PPV$ and $VVP$; $p_1 (p_3)$ and $p_2 (p_4)$ are the momenta of initial (final) state particles; and $t, u$ are two of the Mandelstam variables: $s = (p_1 +p_2)^2, t = (p_1 - p_3)^2,$ and $u = (p_1-p_4)^2$. 

The isospin coefficients $\tau _{I} $  of the reactions listed in     
Eqs.~(\ref{Ampl1}) are determined by considering the charges $Q_{1f}$ 
and $Q_{2f}$ for each of the two particles  in final state, whose
combination gives the total charge $Q = Q_{1f} + Q_{2f} = 0, -1$.   
There are two possible charge configurations $(Q_{1f}, Q_{2f})$ for     
each process in Eq.~(\ref{Ampl1}). The values of $\tau _{I} ^{(i)}$ for
the possible configurations are listed in Table~\ref{tab1}.

\begin{table}[H]
\caption{Isospin coefficients $\tau _{I} $ of the processes described    
  in Eq.~(\ref{Ampl1}) by considering the charges $Q_{1f}$ and $Q_{2f}$ for
  each one of the two particles in final state.}

\centering
\begin{tabular}{cccc}
\hline
\hline
Process & Vertice & $(Q_{1f}, Q_{2f})$ & $\tau_{ij}$\\
\midrule
\multirow{2}{*}{$(a)$} & \multirow{2}{*}{$D^{0,+}D^{*0}\pi^{0,+}$} & (-,0) & $\frac{1}{\sqrt{2}}$ \\
& & (-,+) & $1$ \\
\hline
\multirow{2}{*}{$(b)$} & \multirow{2}{*}{$D^{*0,+} D^{0}\pi^{0,+}$} & (-,0) & $-\frac{1}{\sqrt{2}}$ \\
& & (-,+) & $-1$ \\
\hline
\multirow{2}{*}{ $(c)$ } & \multirow{2}{*}{$D^{*0,+} D^{*0}\pi^{0,+}$} & (-,0) & $\frac{1}{\sqrt{2}}$ \\
& & (-,+) & $1$ \\
\hline
\multirow{2}{*}{ $(d)$ } &  \multirow{2}{*}{$\bar{D}^{0,-}D_{s}^{*-} K^{0,+}$} & (-,0) & $-1$ \\
& & (-,+) & $-1$ \\
\hline
\multirow{2}{*}{ $(e)$ } & \multirow{2}{*}{$\bar{D}^{*0,-}D_{s}^{*-}K^{0,+}$} & (-,0) & $-1$ \\
& & (-,+) & $-1$ \\
\hline
\multirow{2}{*}{ $(f)$ } & \multirow{2}{*}{$\bar{D}^{*0,-} D_{s}^{*-} K^{0,+}$} & (-,0) & $-1$ \\
& & (-,+) & $-1$ \\
\hline
 (g) & $D_{s}^{+}D^{*0}K^{+}$ & (-,+) & $-1$ \\
\hline
(h) & $D_{s}^{*+} D^{*0}K^{+}$ & (-,+) & $-1$ \\
\hline
(i) & $D^{0}D_{s}^{*0}\eta^{0}$ & \multirow{2}{*}{ (-,0)} & $-\frac{\sqrt{6}}{3}$ \\
(j) & $D_{s}^{-} D^{*-} \eta^{0}$ & & $-1$\\
\hline
(k) & $D^{*0}D_{s}^{0}\eta^{0}$ & \multirow{2}{*}{ (-,0)} & $-\frac{\sqrt{6}}{3}$ \\
(l) & $D_{s}^{*-}D_{s}^{-}\eta^{0}$ & & $-1$ \\
\hline
(m) & $D_{s}^{*-} D_{s}^{*-} \eta^{0}$ & (-,0) & $1 $ \\
\hline
\hline
\end{tabular}
\label{tab1}
\end{table}

\section{Cross sections}

\label{Cross Sections}

The isospin-spin-averaged cross section in the center of mass (CM) frame for
the processes in Eq. (\ref{Ampl1}) is given by
\begin{equation}
\sigma_{ab \rightarrow cd} (s) = \frac{1}{64 \pi^{2} s}
\frac{ |\vec{p}_{cd}| }{ |\vec{p}_{ab}|} \int d \Omega \overline{\sum_{S,I}}
| \mathcal{M}_{ab \rightarrow cd} (s, \theta) |^{2} F^{4}, \label{EqCrSec}
\end{equation}
where $\sqrt{s}$ is the CM energy;  $|\vec{p}_{ab}|$ and $|\vec{p}_{cd}|$    
stand for the three-momenta of initial and final particles in the CM frame,
respectively; the symbol $\overline{\sum_{S,I}}$ denotes the sum over the    
spins and isospins of the particles in the initial and final state, weighted 
by the isospin and spin degeneracy factors $g_{1i,r}=(2I_{1i,r}+1)(2I_{2i,r}+1)$ 
and $g_{2i,r}= (2S_{1i,r}+1)(2S_{2i,r}+1)$  of the two particles forming the
initial state, namely: 
\begin{eqnarray}
\nonumber \overline{\sum_{S,I}} |\mathcal{M}_{ab \rightarrow cd} |^{2} &
\equiv   & \frac{1}{g_{a} g_{b}} \sum_{S,I} |\mathcal{M}_{ab \rightarrow cd}
|^{2} \\  
& = & \frac{1}{ g_{a} g_{b} } \sum_{(Q_{1},Q_{2})} \left[ \sum_{S} \vert
\mathcal{M}_{ab \rightarrow cd}^{(Q_{1},Q_{2})} \vert ^{2} \right].
\end{eqnarray}
Finally,  as usual, we have introduced the form factor $F$ to account for
the composite nature of hadrons and their finite extension observed at
increasing momentum transfers. The form factor introduces a suppression of 
the high momentum region and therefore tames the artificial growth
of the cross sections.  
We make use of a monopole-like expression,
defined as~\cite{Hong:2018mpk,Abreu:2021jwm}: 
\begin{eqnarray}
F(\vec{q}) = \frac{ \Lambda^{2} }{ \Lambda^{2} + \vec{q}^{2} },
	  \label{formfactor}
\end{eqnarray}
with $\vec{q}$ being the momentum of the exchanged particle in a $t$- or     
$u$-channel in the center of mass frame, and $\Lambda $ the cut-off,  chosen
to be in 
the range $m_{min} < \Lambda < m_{max}$, taking $m_{min}$ ($m_{max}$)  as the
mass of the lightest (heaviest)  particle entering or exiting the vertices.    
In the present approach we fix $\Lambda = 2.0 $ GeV. For a detailed discussion 
on the role and choice  of the form factor, we refer the reader to
Ref.~\cite{Abreu:2021jwm}.

Using the detailed balance relation,  we can also evaluate the cross sections
of the inverse processes, which lead to the absorption of the $Z_{cs}^{-}$
state.

The calculations of the present work are done with the isospin-averaged masses 
reported in the  PDG~\cite{Zyla:2020zbs}. Since we use a range of values for the 
coupling $g_{Z_{cs}}$ (in order to take into account the uncertainties), the
results are shown in terms of bands.

\begin{figure}[!ht]
    \centering
\includegraphics[{width=1.0\linewidth}]{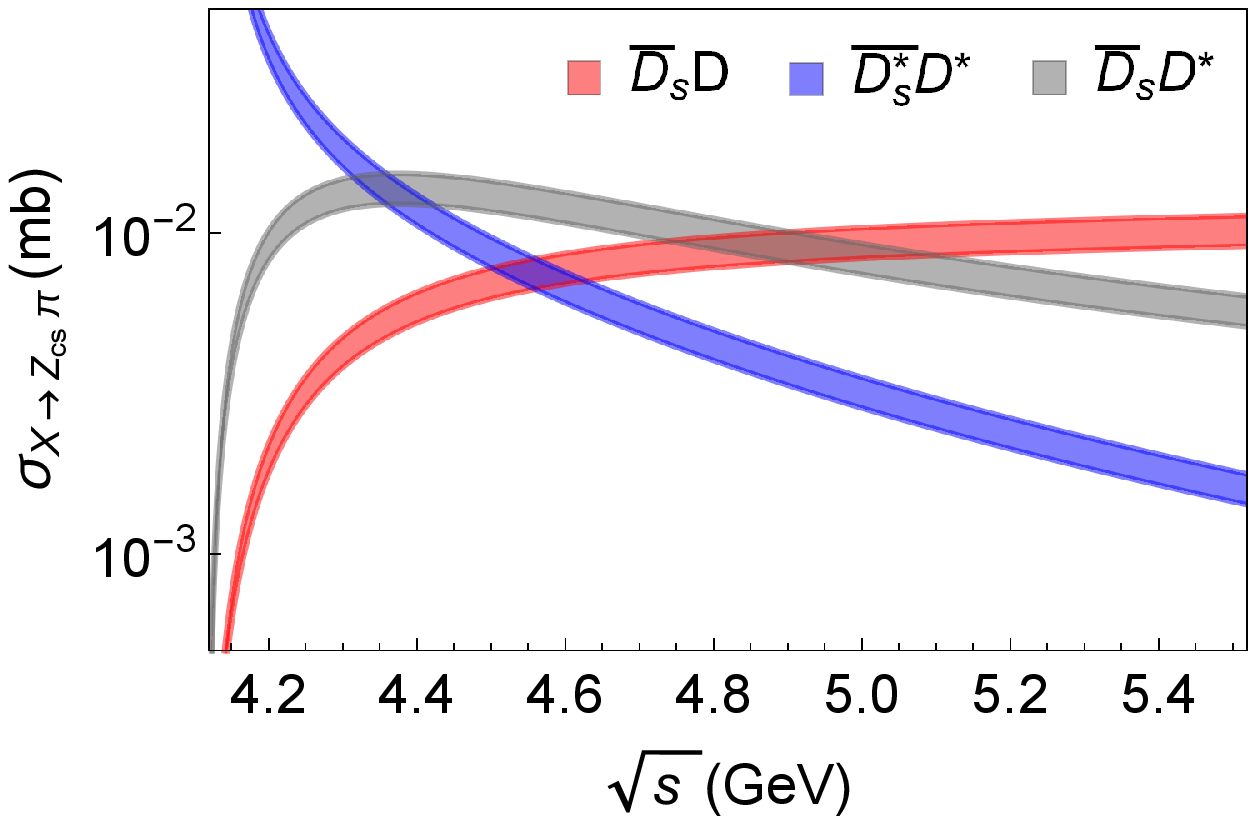}\\
\includegraphics[{width=1.0\linewidth}]{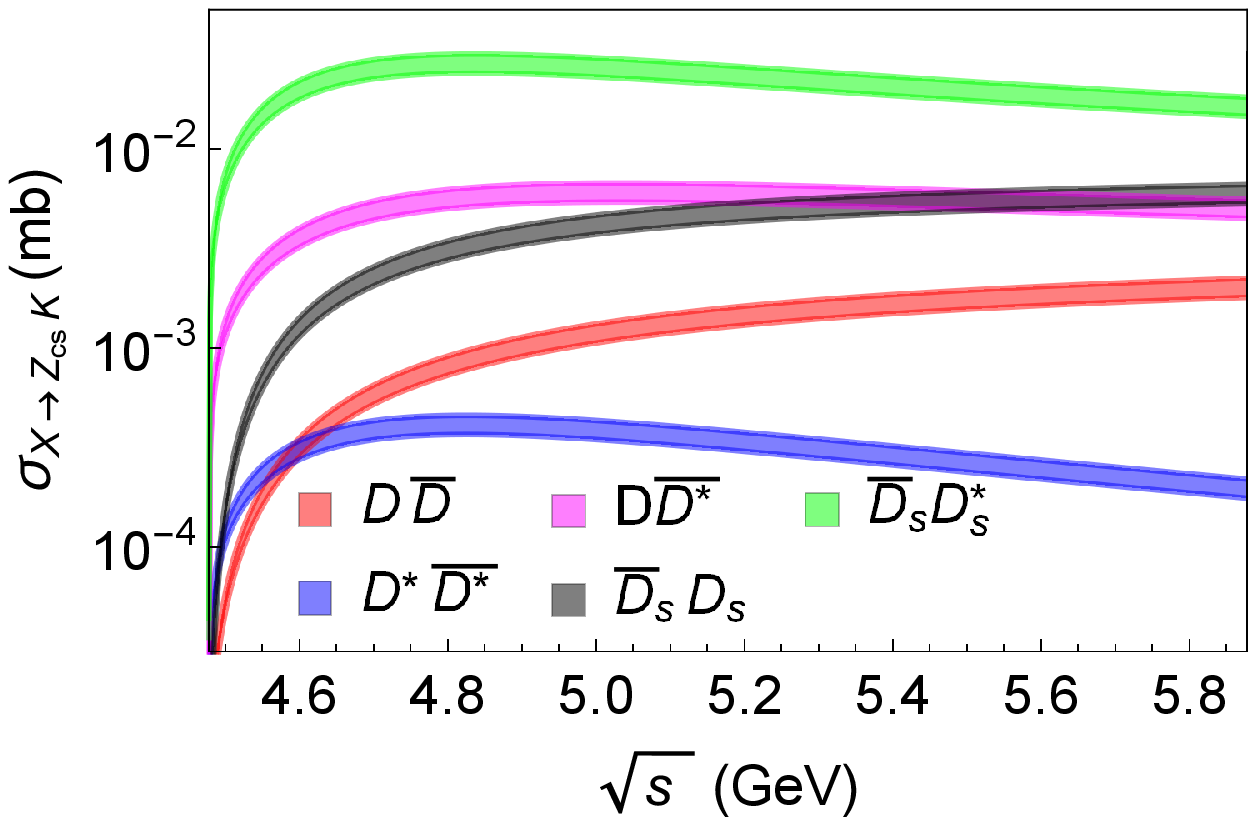}\\
\includegraphics[{width=1.0\linewidth}]{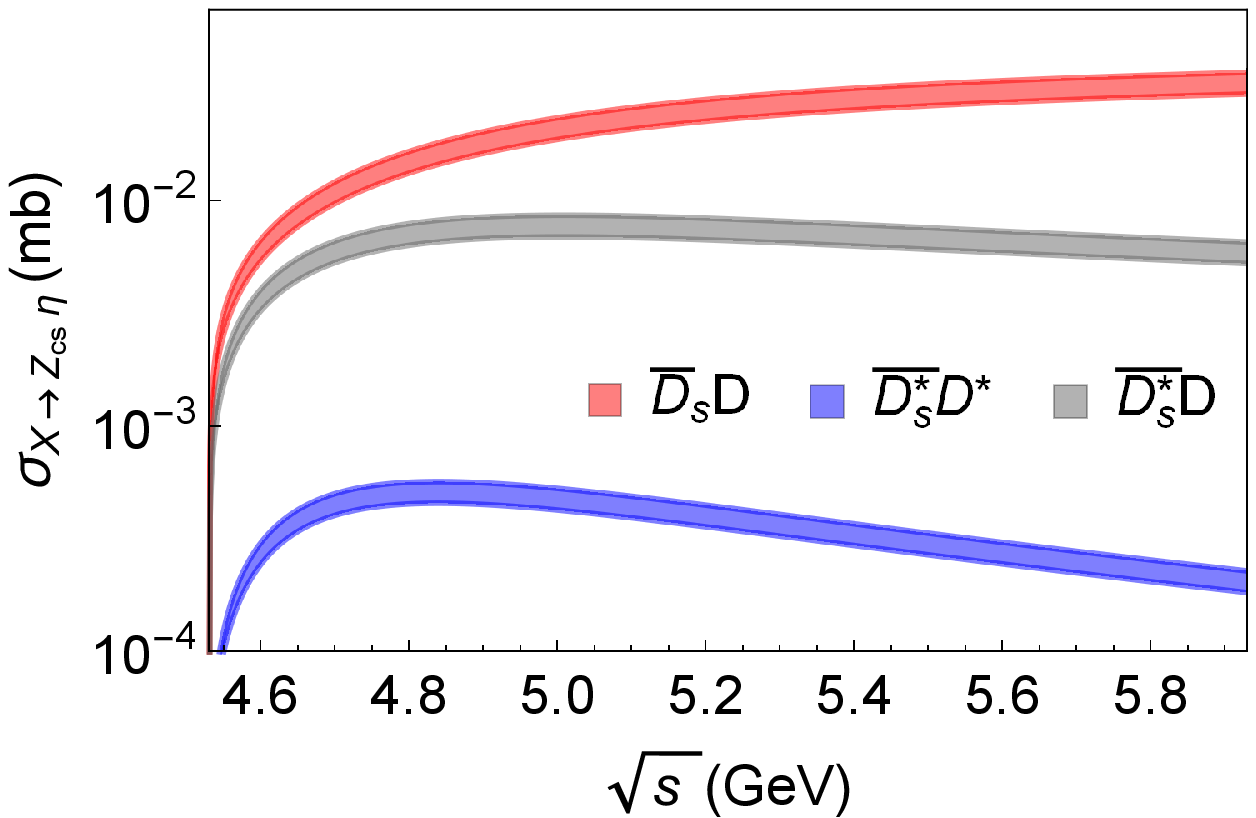}\\
\caption{ Cross sections for the production processes $  Z_{cs}^{-}\pi  $(top),
  $Z_{cs}^{-}K$(center) and $Z_{cs}^{-}\eta$(bottom), as  functions of
  $\sqrt{s}$. }
    \label{Fig:CrSec-Prod}
\end{figure}

The cross sections for the $Z_{cs}^{-}$-production as functions of the CM energy
$\sqrt{s}$  are plotted in Fig.~\ref{Fig:CrSec-Prod}. 
Excluding the contribution of the channel
$D_s ^{*} D^{*} \rightarrow Z_{cs} \pi $, all the cross sections are    
endothermic, having a substantial increase near the threshold and after
that a weak dependence on $\sqrt{s}$.  In the region close 
to the threshold we note that the distinct channels present magnitudes of
the order of $ \sim  10^{-4} -  10^{-2} \mb$. 
For the $Z_{cs}^{-}$ production induced by kaon and $\eta $ mesons, 
the channels with  final states
$D_s \bar{D}_s ^{*}$ and $D_s \bar{D}_s $ have maximal cross sections  at
smaller CM energies. This pattern remains at moderate CM energies
(i.e. $500 \MeV$ above the threshold) for the channels involving
the $Z_{cs} K, \eta $-production,
whereas those of $Z_{cs} \pi $ have closer magnitudes.

\begin{figure}[!ht]
    \centering
              \includegraphics[{width=1.0\linewidth}]{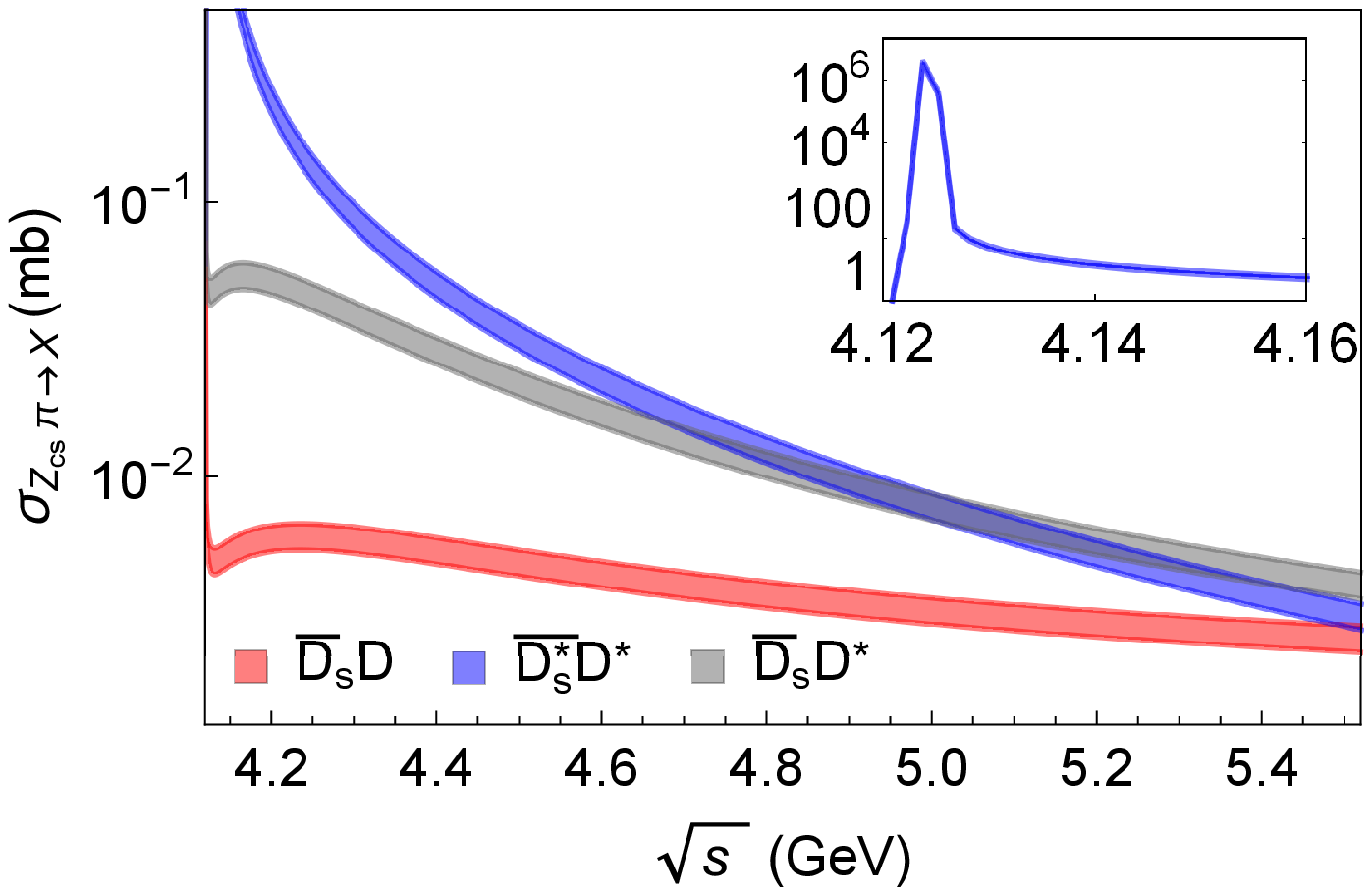}\\
              \includegraphics[{width=1.0\linewidth}]{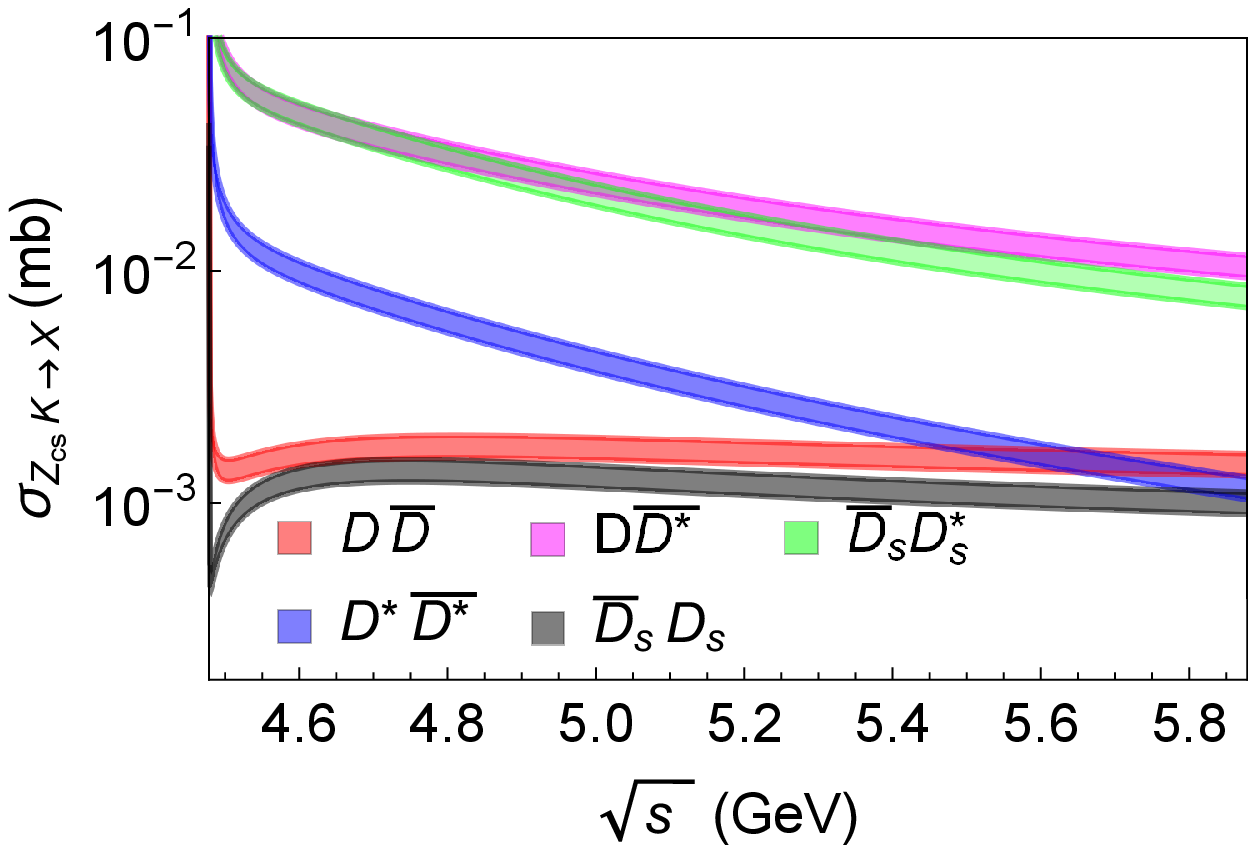}\\
              \includegraphics[{width=1.0\linewidth}]{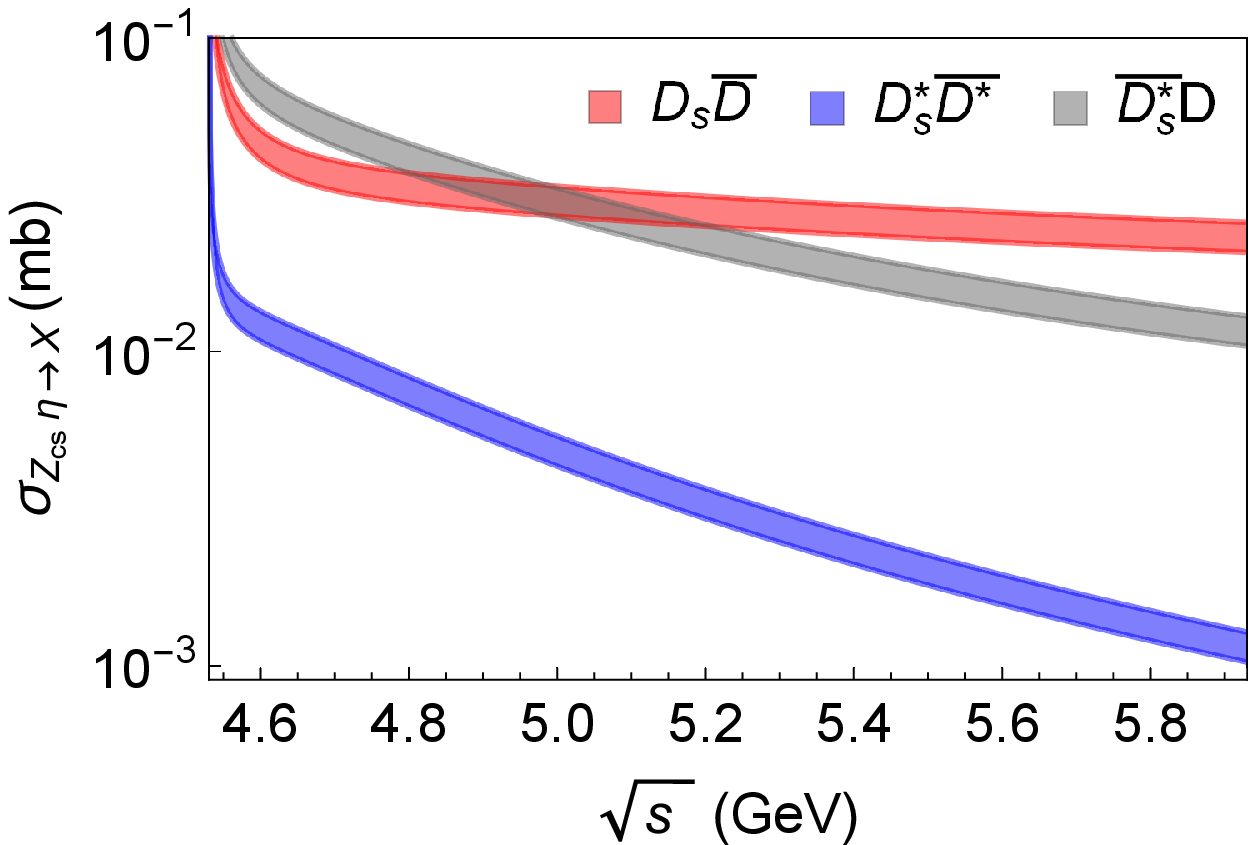}\\
\caption{ Cross sections for the absorption processes $Z_{cs}^{-}\pi$(top), 
  $Z_{cs}^{-}K$(center) and $Z_{cs}^{-}\eta$(bottom), as  
  functions of $\sqrt{s}$. The behavior of the reaction                 
  $ \bar{D}_s ^{*} D^{*} \rightarrow Z_{cs} \pi $ near the threshold is
  evidenced in the inlay panel at the top. }
    \label{Fig:CrSec-Abs}
\end{figure}

Let us now examine the inverse processes. Their cross sections as functions of
the CM energy $\sqrt{s}$  are plotted in Fig.~\ref{Fig:CrSec-Abs}. We see that 
all these absorption cross sections are exothermic, becoming very large near
the threshold. The exception is the case of                            
$ Z_{cs} \pi  \rightarrow \bar{D}_s ^{*} D^{*} $, which has a distinct       
behavior: it starts small at the threshold but rapidly increases and becomes  
very large, and after that decreases as in the other cases. From the region
close 
to the threshold up to moderate energies,  we observe  that the cross sections
are of the order $ \sim  10^{-3} -  10^{-1} \mb$. 

\begin{figure}[!ht]
\centering
\includegraphics[{width=1.0\linewidth}]{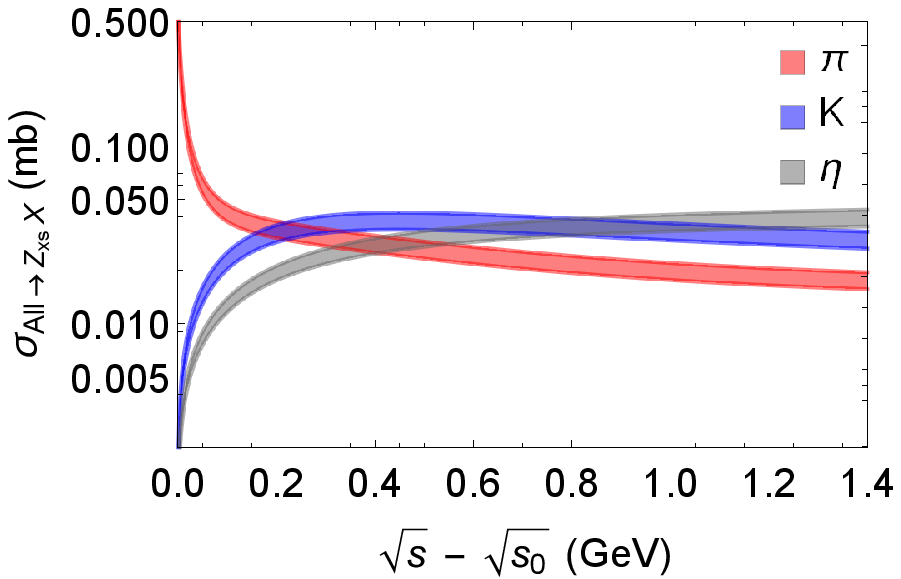}\\
\includegraphics[{width=1.0\linewidth}]{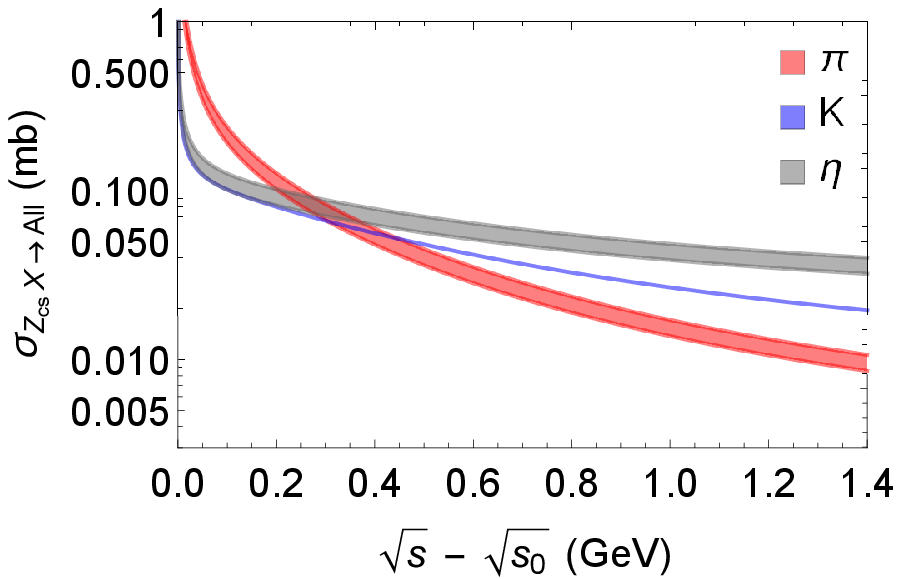}\\    
\caption{Sum of all cross sections for $\mbox{All} \rightarrow Z_{cs}^{-}X$ 
  (top) and $Z_{cs}^{-}X \rightarrow  \mbox{All}$ (bottom), where $X=\pi , K$
  and $\eta$, in function of $\sqrt{s} - \sqrt{s_{0}}$. }
    \label{Fig:CrSec-ProdAbsAll}
\end{figure}

The comparison between $Z_{cs}^{-}$  absorption and production by comoving 
light mesons can be done more easily when the different contributions  are  
added up. The total cross sections for $\mbox{All} \rightarrow Z_{cs}^{-}X$   
and $Z_{cs}^{-}X \rightarrow  \mbox{All}  \, (X=\pi , K, \eta) $ as functions
of $\sqrt{s} - \sqrt{s_{0}}$ ($\sqrt{s_0}$ being the mass threshold for each 
channel) are plotted in Fig.~\ref{Fig:CrSec-ProdAbsAll}. The results suggest   
that the cross sections $\sigma _{ \mbox{All} \rightarrow Z_{cs}^{-} X} $ have 
similar  magnitude and a weak dependence on $\sqrt{s} -\sqrt{s_0} $.
This fact reflects the dynamics as well as the choice of the values of the   
coupling constants.  In the case of absorption processes, this similarity is 
less pronounced and the dependence with $\sqrt{s} -\sqrt{s_0} $ is stronger.

The most important information contained in Fig.~\ref{Fig:CrSec-ProdAbsAll} is
that, for the energy values which are more relevant to
heavy ion collisions ($\sqrt{s} -\sqrt{s_0} < \, 0.6 GeV$), 
$ \sigma_{Z_{cs}^{-}X \rightarrow  \mbox{All}} >
\sigma_{\mbox{All} \rightarrow Z_{cs}^{-}X }$ , i.e. the absorption cross
sections are greater than the  production ones.

In order to better understand this behavior it is useful to rewrite the
ratio of momenta in Eq. (\ref{EqCrSec})  in an expanded and more
instructive form as:
\begin{equation}
\frac{ |\vec{p}_{cd}| }{ |\vec{p}_{ab}|} =
\left(\frac{[s - (m_c + m_d)^2] [ s - (m_c - m_d)^2]}
{[s - (m_a + m_b)^2] [ s - (m_a - m_b)^2]} \right)^{1/2} 
\label{fase}
\end{equation} 
Now let us consider the processes with the largest cross sections:
$Z_{cs} \pi \to \bar{D}_s^* D^*$ and the corresponding inverse process
$\bar{D}_s^* D^* \to  Z_{cs} \pi $. Assuming, just for the sake of the
discussion, that $m_{\pi} = 0$, $m_{\bar{D}_s^*} = m_{{D}^*} = m$  and
$m_{Z_{cs}} = 2m$, and substituting these masses in (\ref{fase}) we find
that the ratio  is $\sqrt{s/(s - 4m^2)}$ for $Z_{cs}$ absorption and it
is $\sqrt{(s - 4m^2)/s}$ for  $Z_{cs}$ production. We see then that the
difference of these two processes comes to a large extent from the phase
space and can be big.

Apart from the ratio of momenta, differences can also be due to the
degeneracy factors. In the absorption process, the initial state is the 
$Z_{cs} - \pi$ system, for which the isospin ($g_I$), spin ($g_S$) and total
($g^a_T$) degeneracy factors are:
\begin{equation}
g^a_T = g_I^Z (=2) \times g_S^Z (= 3) \times g_I^{\pi} (= 3) 
\times g_S^{\pi} (= 1)  = 18 .
\label{gabs}
\end{equation}
For the production process, we have $\bar{D}_s^*$ and $D^*$ in the initial
state and the corresponding degeneracy factors are:
\begin{equation}
g^p_T = g_I^{D^*_s} (=1) \times g_S^{D^*_s} (= 3) \times g_I^{D^*} (= 2)
\times g_S^{D^*} (= 3)  = 18 .
\label{gabs1}
\end{equation}
In this example $g^a_T = g^p_T$ and the difference between absorption and
production comes solely from the phase space. However,  in other process
$g^a_T$ and $g^p_T$ can differ by one order of magnitude.

\section{Thermal cross sections} 

\label{ThermalAvCrossSection} 


Motivated by the results of the previous section, we turn our attention to 
the heavy-ion collision environment. Keeping in mind that the 
temperature of the hadronic medium drives the collision energy, it is
convenient to evaluate the thermal cross sections, defined as convolutions 
of the vacuum cross sections with thermal momentum distributions of the 
colliding particles. This thermal average leads to a strong suppression of
the kinematical configurations very close to the thresholds, and therefore 
threshold effects will not play a relevant role in the presence of a hot
hadronic medium. 

The cross section averaged over the thermal distribution for a reaction 
involving an initial two-particle state going into two final particles
$ab \to cd$ is given by~\cite{ChoLee1,XProd2,Abreu:2017cof,Abreu:2018mnc,Koch}
\begin{eqnarray}
\nonumber \langle \sigma_{ab \rightarrow cd} \upsilon_{ab} \rangle &= &  \frac{\int d^3 \mathbf{p}_a d^3 \mathbf{p}_b f_a (\mathbf{p}_a) f_b (\mathbf{p}_b) \sigma_{ab \rightarrow cd} \upsilon_{ab}}{\int d^3 \mathbf{p}_a d^3 \mathbf{p}_b f_a (\mathbf{p}_a) f_b (\mathbf{p}_b)} \\
\nonumber &= &  \frac{1}{4\beta^2_a K_2 (\beta_a) \beta^2_b K_2 (\beta_b)}\\
\nonumber & & \times \int^{\infty}_{z_0} dz K_1 (z) \sigma (s = z^2 T^2)\\
 & & \times [z^2 - (\beta_a + \beta_b)^2] [z^2 - (\beta_a - \beta_b)^2] \nonumber \\
    \label{AvCrSec}
\end{eqnarray}
where $v_{ab}$ denotes the relative velocity of the two initial interacting    
particles; the function $f_i(\mathbf{p}_i)$ is the Bose-Einstein distribution;
$\beta _i = m_i / T$ ( $T$ being the temperature);
$z_0 = max(\beta_a + \beta_b,\beta_c 
+ \beta_d)$, and $K_1$ and $K_2$ are the modified Bessel functions of
second kind.

\begin{figure}[!ht]
    \centering
              \includegraphics[{width=1.0\linewidth}]{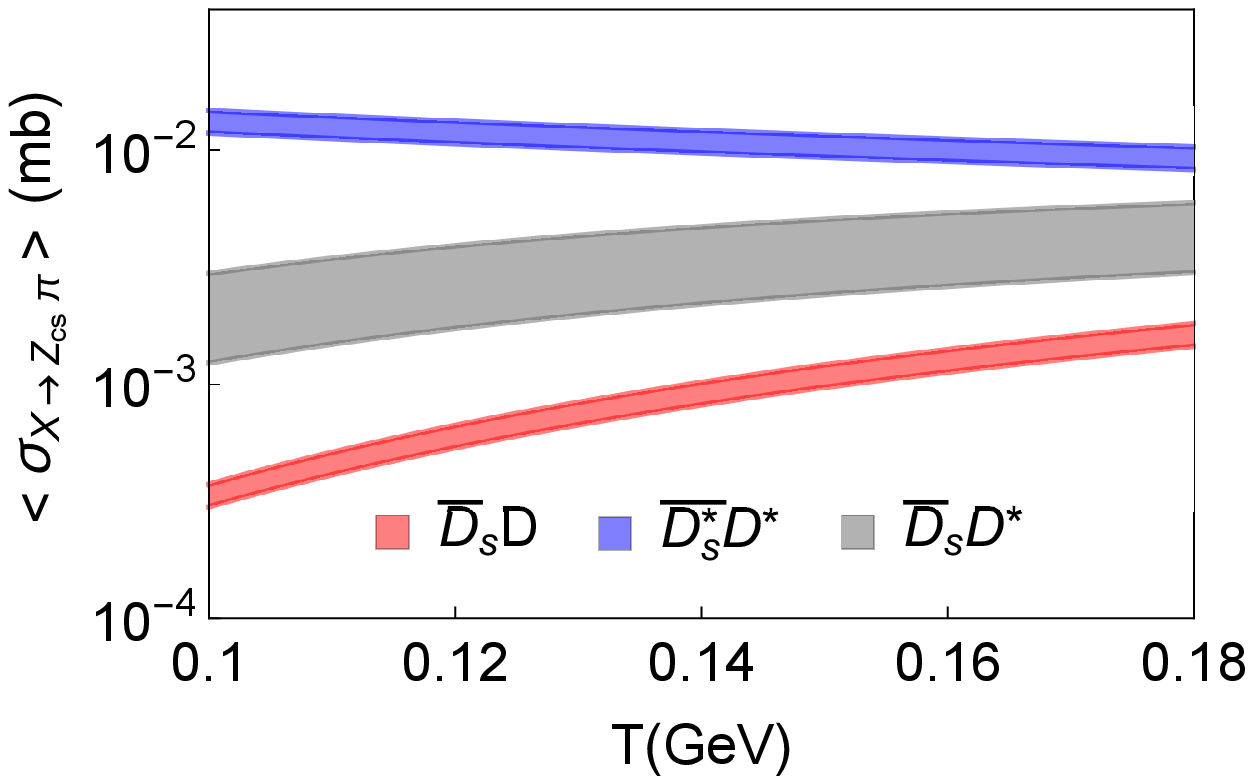}\\
              \includegraphics[{width=1.0\linewidth}]{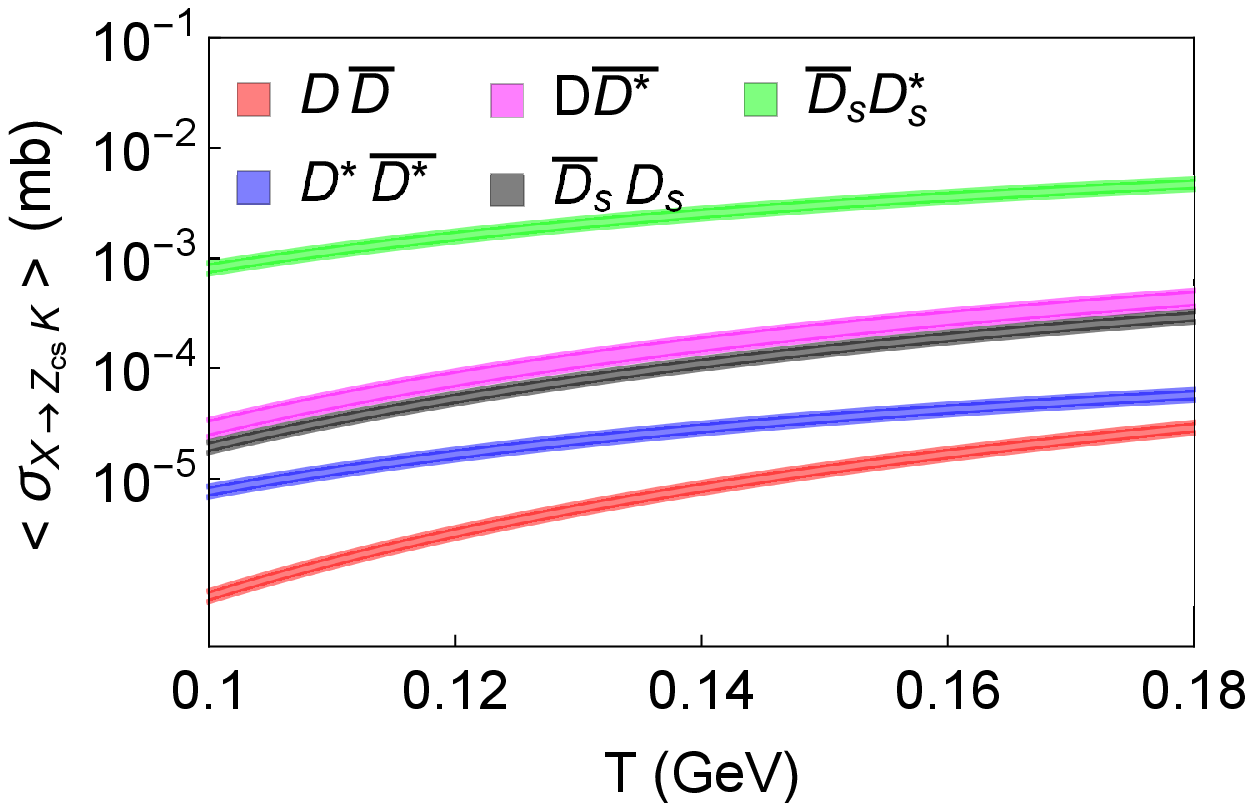}\\
              \includegraphics[{width=1.0\linewidth}]{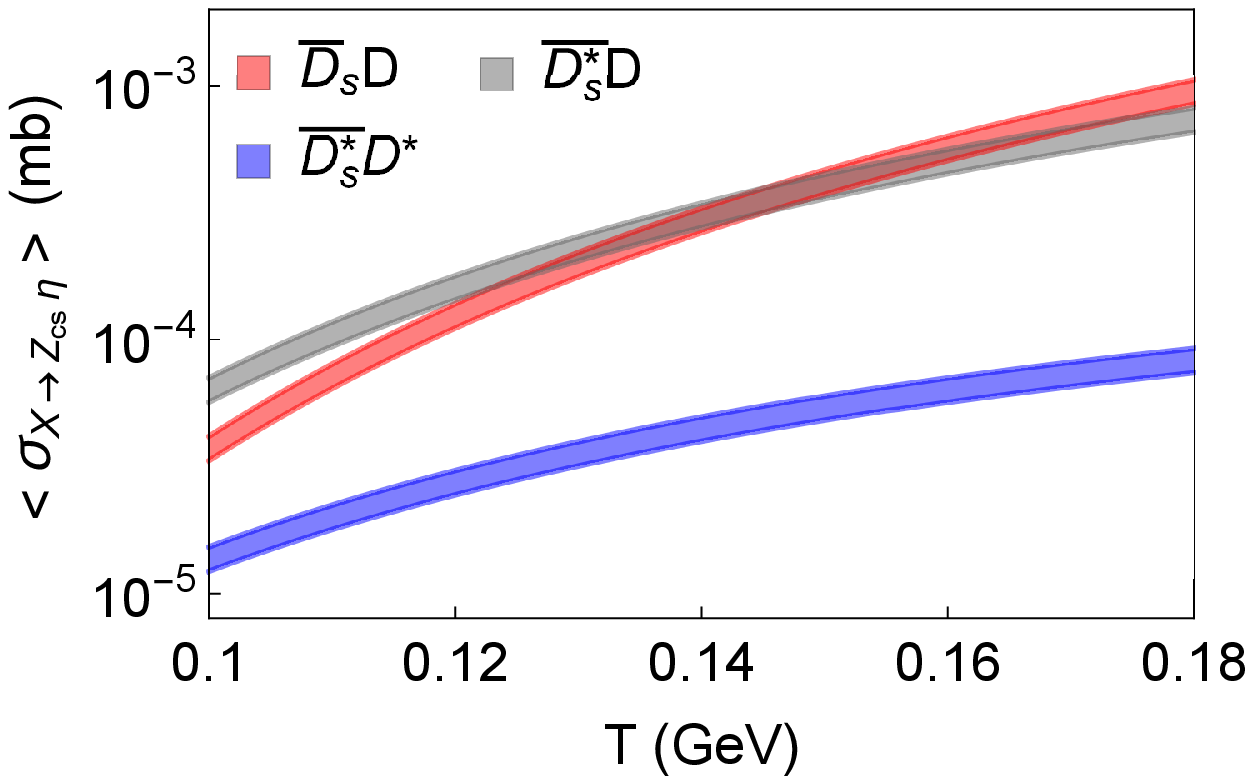}\\
       \caption{Thermal cross sections for the production processes $Z_{cs}^{-}\pi$(top), $Z_{cs}^{-}K$(center) and $Z_{cs}^{-}\eta$(bottom), as a function of temperature $T$. }
    \label{Fig:AvCrSec-Prod}
\end{figure}

\begin{figure}[!ht]
    \centering
              \includegraphics[{width=1.0\linewidth}]{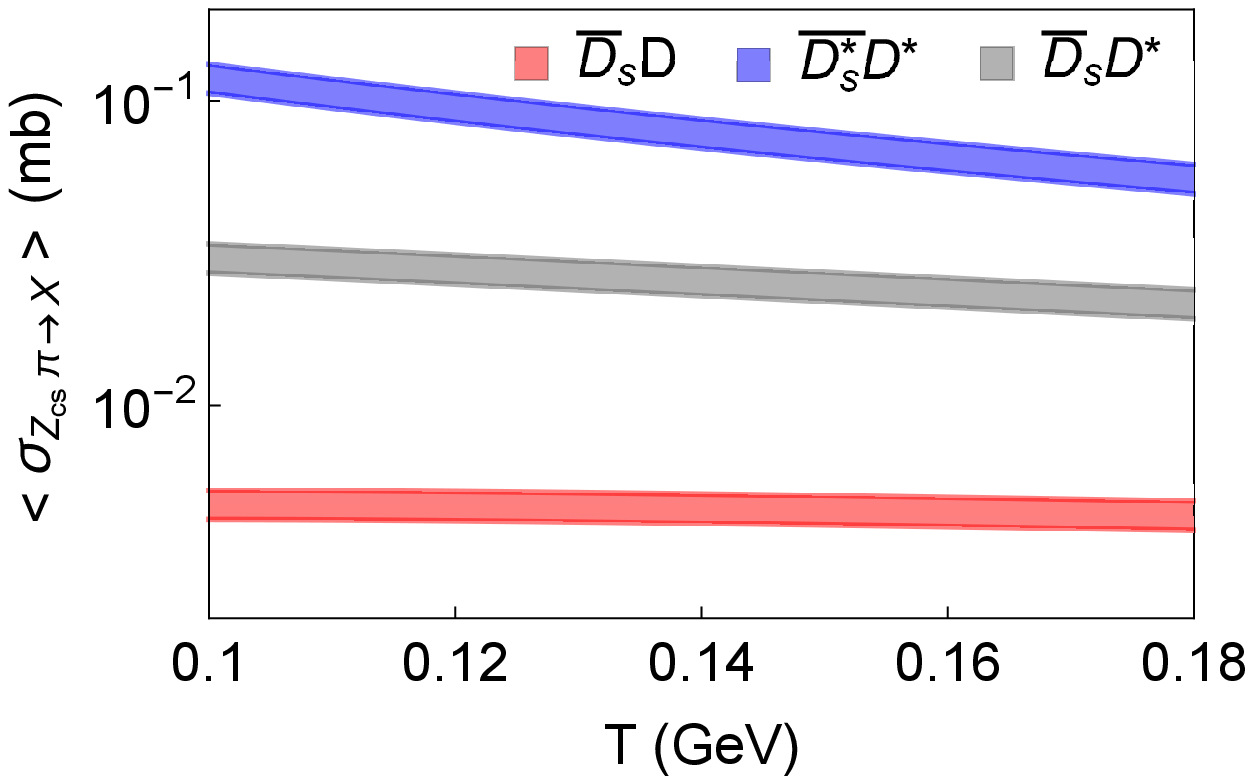}\\
              \includegraphics[{width=1.0\linewidth}]{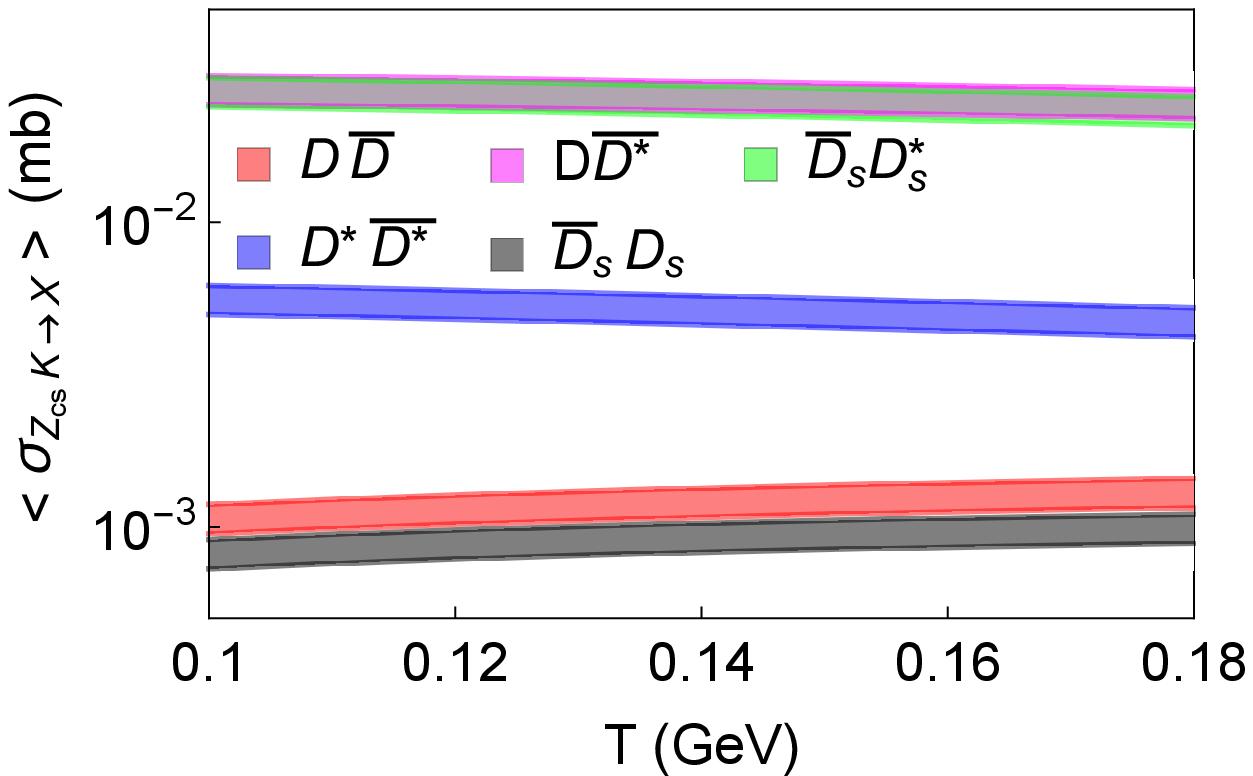}\\
              \includegraphics[{width=1.0\linewidth}]{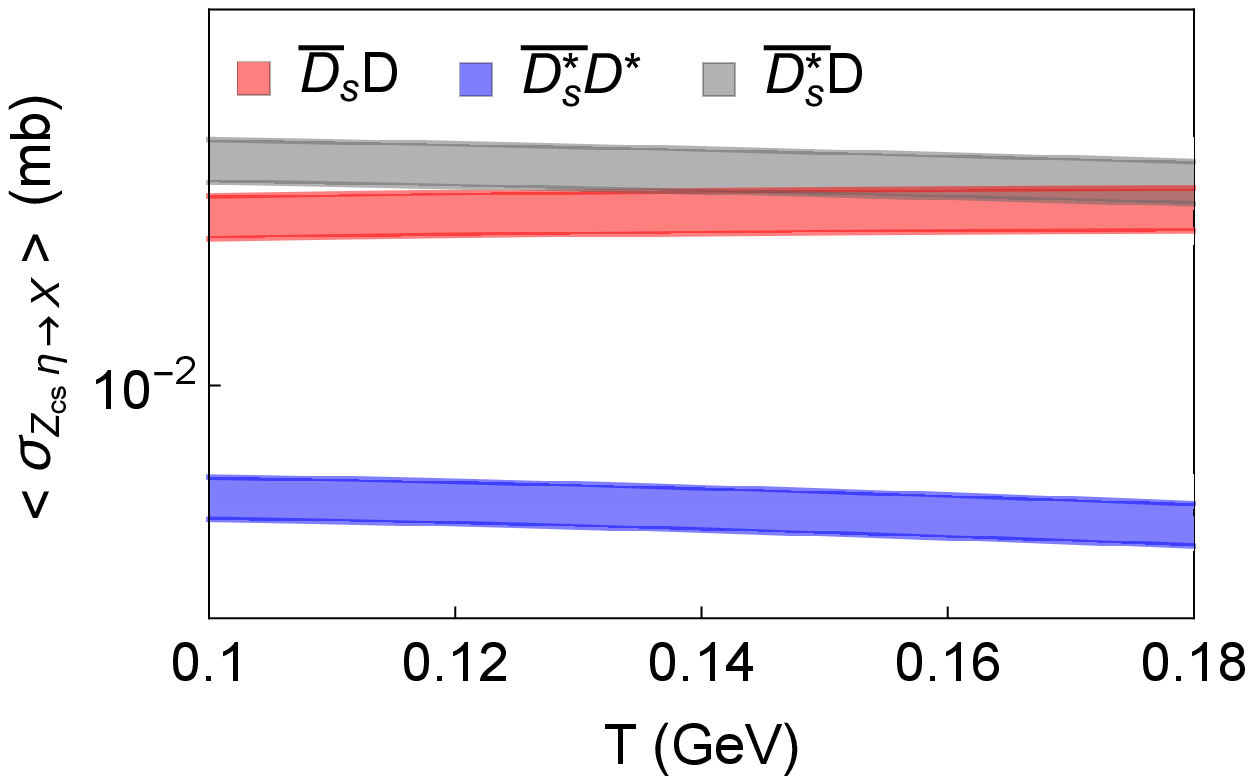}\\
       \caption{Thermal   cross sections for the absorption processes $Z_{cs}^{-}\pi$(top), $Z_{cs}^{-}K$(center) and $Z_{cs}^{-}\eta$(bottom), as a function of temperature $T$. }
    \label{Fig:AvCrSec-Abs}
\end{figure}

In Figs.~\ref{Fig:AvCrSec-Prod} and~\ref{Fig:AvCrSec-Abs} we show the   
thermal  cross sections for $Z_{cs}$ production and absorption plotted as    
functions of the temperature. The results reveal that in general the thermal
cross sections for the $Z_{cs}$ absorption do not change much in 
this range of temperature, staying almost constant. On the other hand, in
the case of $Z_{cs}$ production, most of the cross sections grow
significantly with the temperature. 

These features can be understood from the energy dependence of the cross     
sections shown in Figs.~\ref{Fig:CrSec-Prod} and~\ref{Fig:CrSec-Abs}.
As it can be seen, all the  cross sections (with one exception) of  $Z_{cs}$
production grow with the CM energy and as the temperature increases and the 
charmed mesons in initial state become more energetic (surpassing the        
threshold), the thermal production cross sections grow with $T$.

We emphasize that our most important result is that the thermal cross sections
for  $Z_{cs}$ absorption are greater than those for production, at least     
by one order of magnitude. For instance:  the cross section of
$Z_{cs} \pi  \to  \bar{D}_s ^{*} D^{*} $ is bigger than that for the 
corresponding inverse reaction by one order of magnitude; in the case of the 
channel $ Z_{cs} K \rightarrow D_s  \bar{D}_s $ and its inverse, this difference
is at least of two orders of magnitude, depending on the temperature. 

This result might have important implications for the observed final yield
of the  $Z_{cs}$ state in heavy ion collisions.  The $Z_{cs}$ multiplicity at
the end of the quark-gluon plasma phase (which may be estimated via the
coalescence model) might go through sizeable changes  
because of the interactions during the hadron gas phase. The 
different magnitudes of the thermal cross sections for the
$Z_{cs}$ annihilation and production by comoving hadrons might lead to a
suppression of $Z_{cs}$.



\section{Concluding remarks}

\label{Conclusions}

In this work we have investigated the interactions of the multiquark state 
$Z_{cs}$ with light mesons in the hadron gas phase. We made use of an   
effective Lagrangian framework. The vacuum cross sections as well as the 
thermal  cross sections for the $Z_{cs}X-$ absorption and
production processes ($X=\pi , K, \eta$) have been estimated.

Our results have uncertainties coming from the couplings constants and 
from the form factors (with the corresponding cut-off). Nevertheless, they  
clearly show  that the thermal cross sections for $Z_{cs}$ annihilation are  
larger than the corresponding ones for production. It would be tempting to
conclude that there will be a reduction of the multiplicity of this state    
due to the absorption by the hadron gas. However, in the rate equation which
controls the evolution of the $Z_{cs}$ abundance there are gain and loss     
terms and they depend on the initial number of $D^{(*)}$'s and
$D_s^{(*)}$'s. Since these mesons are much more abundant  
than the  $Z_{cs}$'s, it is not clear a priori what will be the final
outcome. 
A similar feature was also observed in other multiquark states, such as 
the $T_{cc}^+$. In this case, it was observed in \cite{Abreu:2022lfy} 
that the rise or fall of the initial
abundance depended on several factors, including the internal structure
(compact
tetraquark or large meson molecule). This is certainly a very interesting
question and work in this direction is already in progress.

\begin{acknowledgements}

  The authors would like to thank the Brazilian funding agencies for their
  financial support: CNPq (LMA: contracts 309950/2020-1 and 400546/2016-7), 
  FAPESB (LMA: contract INT0007/2016). We are also grateful to the INCT-FNA.

\end{acknowledgements} 



\end{document}